\documentclass[12pt]{article}
\usepackage[margin=1in,footskip=0.25in]{geometry}
\usepackage{graphicx}

\begin{document}
\linespread{1.6}

\title{Remark on Protein Collapse from a Random Walk}

\author{\bf Ramzi Khuri
  \thanks{email: \texttt{ramzi.khuri@baruch.cuny.edu}}}

%\affil{Baruch College, City University of New York\\ Department of Natural Sciences\\ 17 Lexington Avenue, Box A-0506\\ New York, NY 10010, USA}

\maketitle
\center{
Baruch College, City University of New York\\ Department of Natural Sciences\\ 17 Lexington Avenue, Box A-0506\\ New York, NY 10010, USA\\}

\abstract{I use a mean-field approximation to show the rapid collapse of heteropolymers from random walk size to a much smaller, molten globule state due to hydrophobic interactions, to be followed by a slower annealing process in which there is little change in overall size.}

\newpage 
It is known (\cite{HuangLei} and references therein) that a certain class of hydrophobic proteins, when placed in water, undergo a rapid colllapse from the random walk denatured state to a molten globule state, followed by a slow annealing process into the final state. I argue below that this is a generic feature of the random, repulsive interaction due to the water molecules.

Consider a protein-like polymer (more precisely, a heteropolymer with an arbitrary sequence of amino acids, since not all proteins exhibit this behavior) initially in the denatured, random walk state of $n$ steps each
of size $a$, so that the size of the polymer is initially given by 

\begin{equation} \label{randomwalk}
R_0=\sqrt{n} a.
\end{equation}
Suppose we place the (hydrophobic) polymer in a medium of repulsive scatterers, representing the repulsive force due to water molecules, of number density $\rho$ and (dimensionless) potential strength
$u$. The  size of the polymer was shown to be \cite{poly} (see also \cite{polytext}, \cite{HuangLei}, \cite{Echenique} and references therein)
\begin{equation} \label{sizeone}
R^2=x^{-2} \left(1-\exp(-kx^2)\right),
\end{equation}
where $x=u\rho a^2$ can be thought of as an effective scattering cross section and $k = na^2$. Note that (\ref{sizeone}) is essentially a Gaussian interpolation between the random walk and the collapsed polymer: for $kx^2 << 1$, 
\begin{equation} \label{Rsquared}
R^2 \simeq x^{-2} \left(1 - 1 + kx^2 \right) = k,
\end{equation}
from which we recover the random walk result of (\ref{randomwalk}):
\begin{equation} \label{randomwalk}
R \simeq R_0 = \sqrt{k} = \sqrt{n} a,
\end{equation}
where $\sqrt{k}$ is the natural scale size of the random walk polymer. 

For $kx^2 >>1$, 
\begin{equation} \label{collapse} 
R \simeq 1/x << \sqrt{k}
\end{equation}
represents the collapsed polymer.
A more detailed model should still have a Gaussian enveloping function, due to the random placement of the scatterers.

In a mean-field approximation for the collapse process, let $r = R$ represent the radial variable, with the potential $u = u(r)$ depending only on $r$. Since the repulsive net force forces an inward collapse, the effect is that of a mechanical potential $U(r) = - u(r) = -x/(\rho a^2)$, forcing the polymer inward. For simplicity, let us consider a rescaled potential $V(r) = \rho a^2 U(r) = -x(r)$. (\ref{sizeone}) can be rewritten as
\begin{equation} \label{sizetwo}
r = {1\over x} \left(1-\exp(-kx^2)\right)^{1/2},
\end{equation}
where now $x$ is a function of $r$. 
We can plot $V(r)$ by first plotting $r(x)$ from (\ref{sizetwo}), plotting $x(r)$ as the mirror image of $r(x)$ via the diagonal line $r=x$, and then plotting $V(r) = -x(r)$ by symmetry with respect to the $r$-axis. The result is shown in Figure 1 for $k=1$.
\bigskip

\begin{figure}[h]
\includegraphics[scale=.9]{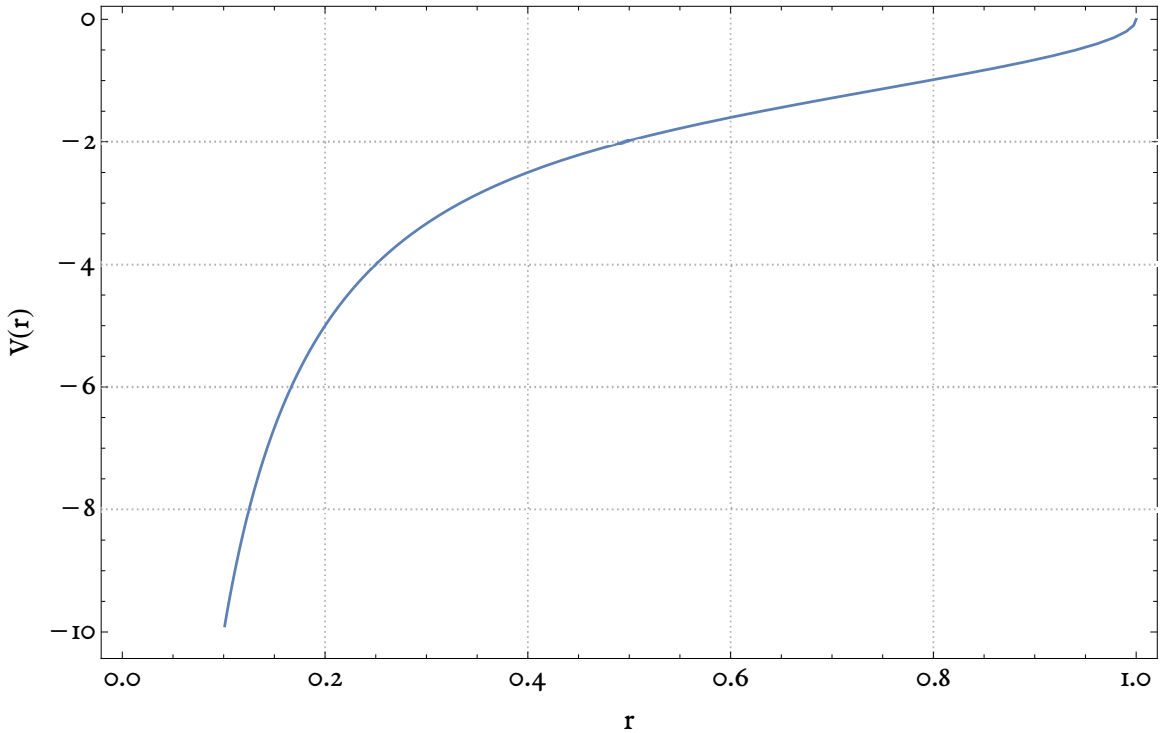}
\caption{Mean-Field Potential}
\end{figure}

 Ideally, I would like to solve for $x(r)$, but can easily show my results by working with $r(x)$. I first calculate $dr/dx$ to be:

\begin{equation} \label{firstderiv}
{dr\over dx} = {-1 + e^{-kx^2} +
 kx^2  e^{-kx^2}\over x^2 \left( 1 -  e^{-kx^2}\right)^{1/2}}
\end{equation}

In the limit $x \to 0$, $r \to \sqrt{k}$ and $dr/dx \to -k^{3/2} x \to 0$. This implies that $dx/dr \to -\infty$, or 
$dV(r)/dr \to +\infty$, implying the rapid collapse of the polymer with the radial force $F(r) \to -\infty$. 
In the limit $x \to \infty$, $dr/dx \to -1/x^2 \to 0$, so again $dV(r)/dr \to + \infty$. 

Of course, the mean-field approximation breaks down well  before the large $x$ domain is reached, but it is not difficult to show that a rapid collapse occurs throughout the domain of validity of this approximation. To simplify the analysis, let $z=r^2/k$ and $y = kx^2$. Then (\ref{sizetwo}) becomes
\begin{equation} \label{sizethree} 
z = {1\over y} \left(1-e^{-y}\right).
\end{equation}
Note that both $z$ and $y$ are dimensionless.
I first need to calculate the derivatives of $z(y)$:

\begin{equation} \label{firstderivdzdy} 
{dz\over dy} = {e^{-y} \left( y + 1 \right) -1 \over y^2}
\end{equation}
and

\begin{equation} \label{secondderiv}
{d^2z\over dy^2} = { -e^{-y} \left(y^2 + 2y + 2\right) +2 \over y^3}.
\end{equation}
A simple exercise in calculus shows that the inflection point equation for the potential  $d^2x/dr^2 = 0$ also solves the equation $d^2r/dx^2 = 0$. This can also be immediately deduced by noting that $r(x)$ and $x(r)$ have the same shape, and inflect for the same values of $r$ and $x$. 
It is straightforward to show that

\begin{equation} \label{drdx} 
{dr\over dx} = {k^2 x \over r} {dz\over dy}.
\end{equation}
The inflection point equation is given by solving the equation

\begin{equation} \label{inflectionrx} {d^2r\over dx^2} = k^2 \left[ {1 \over r} {dz\over dy} - {x\over r^2} {dr\over dx} {dz\over dy} + {x\over r} {d\over dx} \left({dz\over dy}\right) \right] = 0.
\end{equation}
Using $df/dx=2kx df/dy$ and replacing $x$ and $r$ and their derivatives in terms of $y$, $z$ and their derivatives, I arrive at the equation

\begin{equation} \label{inflectionyz} 
\left(1 - {y\over z} \right) {dz\over dy} + 2 y {d^2z\over dy^2} = 0.
\end{equation}
I now replace (\ref{firstderivdzdy}) and (\ref{secondderiv}) into (\ref{inflectionyz}). 
A straightforward but somewhat tedious calculation shows that this is solved for

\begin{equation} \label{inflectionsolution} 
e^y = g(y) =  1 + {y\over 2} + y \sqrt{ \left(y^2 + 1\right) \left( y^2 + 4y +1 \right)}.
\end{equation}
By inspection, (\ref{inflectionsolution}) is solved for $y \simeq 5.36$ up to four significant figures, well beyond the precision of the mean-field approximation. The exponential in (\ref{inflectionsolution}) can be evaluated as

 \begin{equation} \label{inflectionexp} 
e^{5.36} = 212.72 \simeq 212.7
\end{equation}
while the function $g(y)$ on the right hand side  of (\ref{inflectionsolution}) can be evaluated as 

\begin{equation} \label{inflectionpoly}
g(5.36) = 212.74 \simeq 212.7.
\end{equation}
For $y = 5.36$, I get from (\ref{sizethree}) $z \simeq 0.1857$, 
 $dz/dy \simeq -.0338$,
$r \simeq 0.431 \sqrt{k}$, $dr/dx \simeq -0.182 k$ and $dx/dr \simeq -5.49/k$. So even at the point of minimal slope, the collapse rate is relatively fast. Note that it is very important for the physical picture of the polymer collapse that there is only one solution to (\ref{inflectionyz}). This can be seen either by noting that in the plot of the potential in Figure 1 there is only one inflection point, or by noting that in (\ref{inflectionsolution}), the derivative of the exponential function is clearly greater than that of $g(y)$, so that the two functions cannot be equal again for $y > 5.36$.

So what happens past the inflection point, where the mean-field approximation starts to break down? 
The total radial force rapidly becomes much smaller, since the water molecules, upon infliltrating the polymer,  exert their repulsive forces both radially inward and outward. The result is that the net radial force continues to decrease, asympototically approaching zero. This represents the end of the radial collapse, and the much slower formation of the molten globule state. As a result, a slow annealing process can take place.

In summary, I have argued that, for the random, repulsive forces on a protein-like polymer initially in a random walk state, a rapid collapse ensues, followed by a slower annealing process. A more detailed analysis would be interesting, especially one that would take into account the detailed physical properties of the classes of proteins and/or heteroploymers. Hopefully, the above calculation will prove helpful in this regard.

%
%\bigskip
%
%
%
%
%
%\listrefs

\section*{Acknowledgement} I gratefully acknowlege the invaluable help of Bogdan Nicolescu in generating Figure 1 and the proper formatting of the manuscript.

\end{document}